\documentclass[twocolumn]{IEEEtran}

\makeatletter
\let\NAT@parse\undefined
\makeatother

\usepackage{amsmath,amssymb,amsfonts,amsthm}
\usepackage[numbers,sort&compress]{natbib}

\usepackage{mathtools}
\usepackage[nolist,nohyperlinks]{acronym}
\usepackage{color}
\usepackage{multirow}
\usepackage{enumitem}
\usepackage{flushend}

\ifCLASSOPTIONcompsoc
    \usepackage[caption=false, font=normalsize, labelfont=sf, textfont=sf]{subfig}
\else
\usepackage[caption=false, font=footnotesize]{subfig}
\fi

\makeatletter
\def\blfootnote{\xdef\@thefnmark{}\@footnotetext}
\newcases{mycases}{\;}{%
   $\m@th\displaystyle{##}$}{$\m@th\displaystyle{##}$\hfil}{\lbrace}{.}
\makeatother

\newtheorem{proposition}{Proposition}

\begin{acronym}
\acro{PLS}{physical layer security}
\acro{SNR}{signal-to-noise ratio}
\acrodefplural{SNR}[SNRs]{signal-to-noise ratios}
\acro{AWGN}{additive white Gaussian noise}
\acro{OPSC}{outage probability of secrecy capacity}
\acro{CLT}{central limit theorem}
\acro{LoS}{line-of-sight}
\acro{NLoS}{non-line-of-sight}
\acro{LIS}{large-intelligent surfaces}
\acro{CSI}{channel state information}
\acro{PDF}{probability density function}
\acrodefplural{PDF}[PDFs]{probability density functions}
\acro{CDF}{cumulative distribution function}
\acro{TWDP}{two-wave with diffuse power}
\end{acronym}

\begin{document}

\title{On the Beneficial Role of a Finite Number of Scatterers for Wireless Physical Layer Security}
\author{Pablo Ram\'irez-Espinosa, R. Jos\'e S\'anchez-Alarc\'on and F. Javier L\'opez-Mart\'inez}

\maketitle

\blfootnote{\noindent Manuscript received November xx, 2019; revised XXX. This work has been funded by the Spanish Government and the European Fund for Regional Development FEDER (project TEC2017-87913-R). The review of this paper was coordinated by XXXX.}

\blfootnote{\noindent The authors are with Departmento de Ingenieria 
de Comunicaciones, Universidad de Malaga - Campus de Excelencia Internacional Andalucia Tech., Malaga 29071, Spain (e-mail: $\rm \{pre, rjsa, fjlopezm\}@ic.uma.es$).}

\blfootnote{Digital Object Identifier 10.1109/XXX.2019.XXXXXXX}

\blfootnote{This work has been submitted to the IEEE for publication. Copyright may be transferred without notice, after which this version may no longer be accessible.}

\begin{abstract}
We show that for a legitimate communication under multipath quasi-static fading with a reduced number of scatterers, it is possible to achieve perfect secrecy even in the presence of a passive eavesdropper for which no channel state information is available. Specifically, we show that the outage probability of secrecy capacity (OPSC) is zero for a given range of average signal-to-noise ratios (SNRs) at the legitimate and eavesdropper's receivers. As an application example, we analyze the OPSC for the case of two scatterers, explicitly deriving the relationship between the average SNRs, the secrecy rate $R_S$ and the fading model parameters required for achieving perfect secrecy. The impact of  increasing the number of scatterers is also analyzed, showing that it is always possible to achieve perfect secrecy in this scenario, provided that the dominant specular component for the legitimate channel is sufficiently large compared to the remaining scattered waves.
\end{abstract}


\section{Introduction}

In the last decade, the seminal works in \cite{Bloch2008,Liang2008,Gopala2008} have boosted the interest of the research community for providing secure communications over wireless channels from an information-theoretic viewpoint based on the classical work by Shannon \cite{Shannon1949}. Compared to the case on which fading is neglected \cite{Wyner1975,Leung1978}, the effect of random fluctuations due to fading turns out being beneficial in many cases. For instance, secure communications are possible even when the legitimate receiver experiences a lower average \ac{SNR} than the eavesdropper; in other circumstances, the secrecy capacity under fading may be larger than its \ac{AWGN} counterpart. In those cases on which channel state information of the eavesdropper is unknown at the legitimate transmitter, these previous works \cite{Bloch2008,Liang2008,Gopala2008} show that it is not possible to ensure \emph{perfect secrecy}, and only a probabilistic measure of secrecy is available through the \ac{OPSC} \cite{Bloch2008}.

The field of wireless \ac{PLS} has now become a rather mature field and numerous works have been devoted to characterize the key performance metrics under different propagation conditions \cite{Wang2014,Lei2017,Zeng2018}. Classically, state-of-the-art fading models like those considered in \cite{Bloch2008,Liang2008,Gopala2008} are based on the \ac{CLT} assumption. This gives rise to the Rician and Rayleigh models, or generalizations of these \cite{Durgin2002,Yacoub07,Romero2017}. The presence of a diffusely propagating component arising from the superposition of a sufficiently large number of non-dominant received waves is common to all these models. In a way, the \ac{CLT} provides an approximate solution to the sum of random phase vectors, which is one of the key problems in communication theory \cite{Slack1946,Rice1974,Simon1985}.

Nowadays, because of the new use cases of wireless systems under the umbrella of 5G and its evolutions, there are several examples on which the propagation conditions may be substantially different to those predicted by state-of-the-art fading models. For instance, in mmWave communications a scarce number of multipath components arrives at the receiver \cite{Niu2015}, so that diffuse scattering only becomes relevant when \ac{NLoS} conditions are considered \cite{Solomitckii2016}. In a different context, the potential of \ac{LIS} \cite{Subrt2012} to design the amplitude and phases of the scattered waves in order to optimize system performance can also be translated into a superposition of a finite number of individual waves.

Due to the great deal of attention received by these aforementioned emerging scenarios, we revisit in this work the issue of secure communications over wireless channels, with one key question in mind: \emph{What's the effect of considering a finite number of scatterers on wireless physical layer security?} For the first time in the literature, we demonstrate that it is possible to achieve \emph{perfect secrecy} for the communication between two legitimate peers under multipath quasi-static fading, i.e, \emph{zero} \ac{OPSC}, as the number of scatterers is reduced. We determine the conditions on which perfect secrecy can be ensured, and then we give some practical examples using a ray-based fading model with an increasing number of scattered waves. We also observe that using the alternative definition of \ac{OPSC} in \cite{Zhou2011}, which, in contrast to those in \cite{Bloch2008,Liang2008,Gopala2008}, only accounts for outage events that actually compromise the security of the communication, secrecy performance can be further improved. 

The remainder of this paper is structured as follows. The system model for \ac{PLS} over fading channels with a finite number of rays is formulated in Section \ref{S2}. Then, the notion of perfect secrecy over ray-based fading channels is introduced in Section \ref{S3}. As an illustrative example, the two-ray case is analyzed in Sections \ref{S4} and \ref{S5}, showing how secure and reliable transmission can be attained. The effect of considering a larger number of rays is analyzed in Section \ref{S6}, whereas main conclusions are drawn in Section \ref{S7}.

\section{Problem Formulation}
\label{S2}
\subsection{System model for \ac{PLS}}
We consider a legitimate user (Alice) who wants to send confidential messages to another user (Bob) in the presence of an eavesdropper (Eve). For simplicity, yet without loss of generality, all these agents are equipped with single-antenna devices. The complex channel gains from Alice to Bob and Eve are denoted by $h_b$ and $h_e$, respectively, and assumed constant during the transmission of an entire codeword but independent from one codeword to the next one, i.e. we consider quasi-static fading channels. Therefore, the instantaneous \acp{SNR} at Bob and Eve are given by
\begin{align}
	\gamma_b &= \overline{\gamma}_b \frac{\|h_b\|^2}{\mathbb{E}[\|h_b^2\|]}, &  \gamma_e &=  \overline{\gamma}_e \frac{\|h_e\|^2}{\mathbb{E}[\|h_e^2\|]},
\end{align}
where $\mathbb{E}[\cdot]$ is the expectation operator and $\overline{\gamma}_b$ and $\overline{\gamma}_e$ denote the average \ac{SNR} at Bob and Eve, respectively. 



If Alice has perfect knowledge of both Bob's and Eve's instantaneous \ac{CSI}, perfect secrecy can be obtained by adapting the transmision rate, $R_s$, in those instants where $\gamma_b > \gamma_e$ \cite{Bloch2008,Gopala2008}. The secrecy capacity, i.e., the maximum transmission rate ensuring a secure communication between Alice and Bob, is obtained by leveraging the classical results over real Gaussian channels in \cite{Wyner1975, Leung1978} to model complex ones, leading to\footnote{Unless specifically stated, all the logarithmic functions in this paper are base $2$.} 
\begin{equation}
	\label{eq:CsDef}
	C_s = [C_b - C_e]^+ = [{\rm log}(1+\gamma_b) - {\rm log}(1+\gamma_e)]^+,	
\end{equation} 
where $C_b$ and $C_e$ are the capacities of Bob and Eve, respectively, and $[x]^+$ is the shorthand notation for ${\rm max}\{0,x\}$ . Thus, for each channel realization, Alice would transmit at a rate $R_s \leq C_s$ 
 in order to avoid any information leakage to Eve. 

Consider now a more realistic case in which Eve's instantaneous \ac{CSI} is unknown at the transmitter (corresponding, e.g., to that of a purely passive eavesdropper). In this case, previous works state that perfect secrecy cannot be achieved, and therefore they resort on outage analysis \cite{Bloch2008,Liang2008,Gopala2008,Zhou2011}. That is, Alice would blindly establish a target transmission rate, $R_s$, relying on the assumption that the secrecy capacity of the channel is larger than $R_s$. If $C_s < R_s$, then an outage occurs and the security of the tranmission is compromised with some probability. The interest lies then in the analysis of the probability of such event, namely \ac{OPSC}, and defined as \cite{Bloch2008, Zhou2011}
\begin{equation}
	\label{eq:PoutDef}
	P_{\rm out}(R_s) \triangleq P\{C_s < R_s\}.
\end{equation}  

However, all the aforementioned works consider that channel gains and, consequently, $\gamma_b$ and $\gamma_e$, are distributed according to classical fading models arising from the assumption of \ac{CLT}, which may not be suitable to characterize channel conditions in some emerging scenarios such as mmWave communications or propagation through \ac{LIS} \cite{Niu2015, Solomitckii2016}. Because of the capital importance of the \acp{SNR} distributions in the outage analysis, in the next section we investigate the impact of ray-based models in the \ac{OPSC}, proving that it is possible to achieve perfect secrecy even when Eve's \ac{CSI} remains unknown at the transmitter. 

\subsection{\ac{CLT} and ray-based fading models}
Due to the multipath propagation, the complex based-band received signal is written as the superposition of multiple waves arising from reflections and scattering as \cite[eq. (1)]{Durgin2002}
\begin{equation}
	\label{eq:NrayPhysical}
	h = \sum_{i=1}^N V_i e^{j\phi_i}
\end{equation}
where $N$ denotes the number of multipath waves, $V_i\in\mathbb{R}^+$ their constant amplitudes and $\phi$ their phases, which are assumed to be statistically independent and uniformly distributed over $[0,2\pi)$. Traditionally, the sum in \eqref{eq:NrayPhysical} is split into two groups of waves as
\begin{equation}
	h = \sum_{i=1}^M V_i e^{j\phi_i} + \sum_{i=1}^P \widehat{V}_i e^{j\theta_i}
\end{equation} 
where $\theta_i$ $\forall$ $i$ are also independent and uniformly distributed. Hence, the first sum represents the contribution of the $M$ dominant or specular components, whilst the second one groups the contribution of non-specular or diffuse waves, where the power of each component is considerably lower. Thus, the dominant waves are associated with the \ac{LoS} components, whereas the diffuse part represents the contribution of reflections and scattering. When $P$ is sufficiently large, i.e., we have a rich multipath propagation, the diffuse component can be regarded as Gaussian because of the \ac{CLT}, and therefore
\begin{equation}
	\label{eq:CLTmodelsPhysical}
	h = \sum_{i=1}^M V_i e^{j\phi_i} + \sigma_x X + j\sigma_y Y
\end{equation} 
with $X,Y\sim\mathcal{N}(0,1)$.

Equation \eqref{eq:CLTmodelsPhysical} is the basis for most popular fading models, and the widespread classical distributions arise depending on the value of the parameters $M$, $\sigma_x$ and $\sigma_y$. For instance, if $\sigma_x = \sigma_y$ and $M = 0$ we obtain the Rayleigh model, whilst $M=1$ yields the Rice distribution and $M=2$ reduces to the \ac{TWDP} distribution \cite{Durgin2002}. In this work, we will stick to the general formulation in \eqref{eq:NrayPhysical} in order to explicitly account for the effect of considering a finite number of multipath waves on \ac{PLS}.
%
%
%
\section{Perfect secrecy over fading channels}
\label{S3}
\subsection{Impact of a reduced number of scatterers in \ac{OPSC}}
In order to better understand the influence of the fading distribution in the \ac{OPSC}, we reformulate $P_{\rm out}$ in \eqref{eq:PoutDef} in terms of Bob's and Eve's \acp{SNR} as
\begin{equation}
	\label{eq:PoutGameq}
	P_{\rm{out}}(R_s) = P\{\gamma_b < 2^{R_s}\gamma_e +2^{R_s}-1\} = P\{\gamma_b < \gamma_{\rm eq}\},
\end{equation}
which is obtained by introducing \eqref{eq:CsDef} in \eqref{eq:PoutDef} and performing some basic algebraic manipulations. Note that, when conditioning on $\gamma_e$, $P_{\rm out}$ corresponds to the \ac{CDF} of $\gamma_b$ and, therefore, it can be computed by averaging over all the possible states of $\gamma_e$ as 
\begin{equation}
	\label{eq:PoutIntegral}
	P_{\rm out}(R_s) = \int_0^\infty F_{\gamma_b}\left(2^{R_s}\gamma_e + 2^{R_s}-1\right)f_{\gamma_e}(\gamma_e)\,d\gamma_e.
\end{equation}

Regarding \eqref{eq:PoutGameq}, it is clear that the condition for  secrecy is $\gamma_b > \gamma_{\rm eq}$, where $\gamma_{\rm eq}$ is obtained from $\gamma_e$ as $\gamma_{\rm eq} =2^{R_s}\gamma_e +2^{R_s}-1$. Geometrically, the outage probability is therefore given by the common area under the \acp{PDF} of $\gamma_b$ and $\gamma_{\rm eq}$, being the latter a reescaled and shifted version of $f_{\gamma_e}(\gamma_e)$ of the form 
\begin{equation}
	f_{\gamma_{\rm eq}}(\gamma_{\rm eq}) = 2^{-R_s} f_{\gamma_e}(2^{-R_s}(\gamma_{\rm eq}+1)-1).
\end{equation}

\begin{figure}[t]
\centering
\subfloat[$\gamma_b$ and $\gamma_{\rm eq}$ follow a fading distribution (Rician) arising from \ac{CLT}. \label{fig:1a}]{\includegraphics[width=1.0\columnwidth]{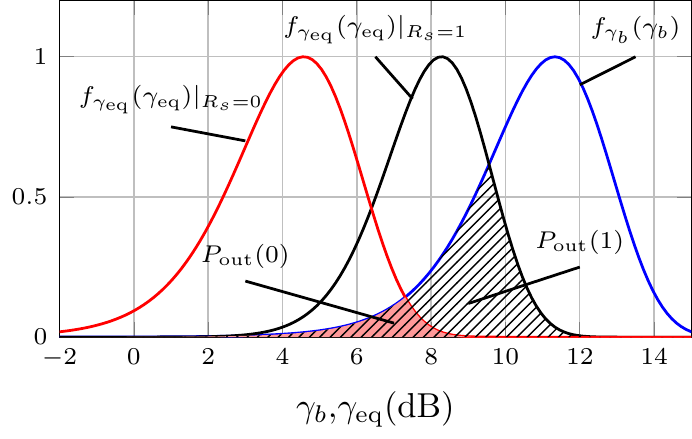}} 
\\
\subfloat[$\gamma_b$ and $\gamma_{\rm eq}$ follow a ray-based fading distribution. \label{fig:1b}]{\includegraphics[width=1.0\columnwidth]{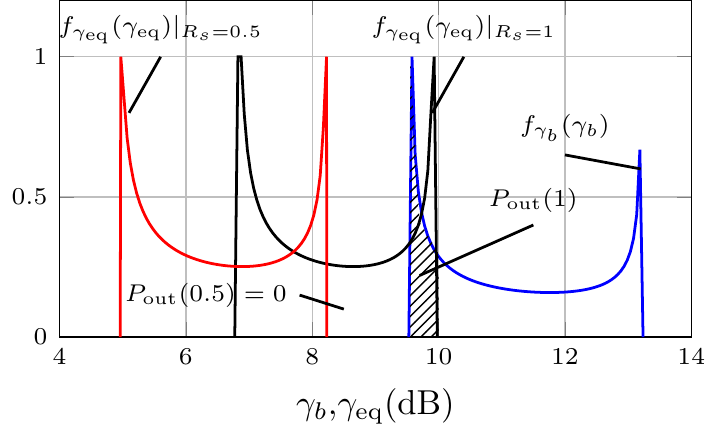}}
\caption{Graphical definition of the \ac{OPSC} as the common area under the \acp{PDF} of $\gamma_b$ and $\gamma_{\rm eq}$ for classical and ray-based fading models, and different values of $R_s$ ($\overline{\gamma_b} = 12 \;{\rm dB}$ and $\overline{\gamma_e} = 5 {\rm \;dB}$). For better visualization, the \acp{PDF} in the figure have been normalized.}
\label{fig:1}
\end{figure}

Thus, if we consider any fading distribution arising from the \ac{CLT} assumption as in \eqref{eq:CLTmodelsPhysical}, the \acp{PDF} of the \acp{SNR} --  or, equivalently, those of $\|h\|^2$ -- are supported on a semi-infinite interval, and then the tails of  $f_{\gamma_b}(\gamma_b)$ and $f_{\gamma_{\rm eq}}(\gamma_{\rm eq})$ overlap regardless of the values of $R_s$ and the average \acp{SNR}. Hence, the condition of $\gamma_{b} < \gamma_{\rm eq}$ is met with non-null probability and perfect secrecy cannot be achieved, as stated in \cite{Bloch2008,Liang2008,Gopala2008,Zhou2011}. This can be observed in Fig. \ref{fig:1a}, where even for $R_s = 0$ there exists some outage area. 

However, things are different when assuming ray-based fading models. Due to the consideration of a finite number of waves, there is a maximum and a minimum value for both the channel gains and the instantaneous \acp{SNR}, i.e., the \acp{PDF} of $\gamma_b$ and $\gamma_{\rm eq}$ are supported on a bounded interval, say $[\gamma^{\rm min}, \gamma^{\rm max}]$. These limit values will depend on the relative amplitudes of the incident waves, that will add-up destructively/constructively with some probability. Therefore, it is evident that in some cases the distribution domains will be disjoint, and hence the \ac{OPSC} can be identically zero, as showed in Fig. \ref{fig:1b}. This is an important observation, since it will allow us to achieve perfect secrecy for transmission rates $R_s > 0$ without Eve's \ac{CSI} knowledge at the transmitter. In the next subsection we formalize this observation and give the condition to achieve perfect secrecy over ray-based fading channels.

\subsection{Achieving perfect secrecy over ray-based fading channels}

Let us consider that the gains for both Eve's and Bob's channels are given by \eqref{eq:NrayPhysical}, and assume, without loss of generality, that $V_i \geq V_j$ $\forall$ $i<j$. It is clear that the maximum value of $h_k$, with $k = b,e$ denoting indistinctly Bob's and Eve's gains, is obtained when all the waves in \eqref{eq:NrayPhysical} are summed coherently. In turn, the minimum value arises when destructive combination occurs. Consequently, and in stark contrast with classical fading distributions, the domain of $\|h_k\|$ is bounded on the interval $[\|h_k^{\rm min}\|, \|h_k^{\rm max}\|]$ with
\begin{align}
	\label{eq:GainsMinMax}
	\|h_k^{\rm min}\| &= \left[V_{1,k} - \sum_{i=2}^{N_k}V_{i,k}\right]^+, &  \|h_k^{\rm max}\| &= \sum_{i=1}^{N_k} V_{i,k}.
\end{align}  

 Therefore, this finite domain definition of channel gains allows us to achieve zero \ac{OPSC} when a certain condition is met, as stated in the following proposition. 


\begin{proposition}
	\label{prop:1}
	Consider $h_b$ and $h_e$ as in \eqref{eq:NrayPhysical}. Then, for a given transmission rate $R_s >0$, perfect secrecy, i.e., $P_{\rm out}(R_s) = 0$, is achieved if
	\begin{equation}
		\label{eq:PerfSecCondition}
		\gamma_b^{\rm min} > 2^{R_s} \gamma_e^{\max} + 2^{R_s} -1,
	\end{equation}
	where $\gamma_b^{\rm min}$ and $\gamma_e^{\rm max}$ are given by 
	\begin{align}
	\label{eq:SNRMinMax}
	\gamma_b^{\rm min} &= \overline{\gamma}_b\frac{\|h_b^{\rm min}\|^2}{\mathbb{E}[\|h_b\|^2]}, & \gamma_e^{\rm max} = \overline{\gamma}_e\frac{\|h_e^{\rm max}\|^2}{\mathbb{E}[\|h_e\|^2]}
\end{align}	
	with $\|h_b^{\rm min}\|$ and $\|h_e^{\rm max}\|$ as in \eqref{eq:GainsMinMax} and 
	\begin{equation}
		\label{eq:ExpectationNrayPower}
		\mathbb{E}[\|h_k\|^2] = \sum_{i=1}^{N_k} V_{i,k}^2, \quad k=b,e.
	\end{equation}
\end{proposition}
\begin{IEEEproof}
	The condition for zero \ac{OPSC} is given by $\gamma_b^{\rm min}>\gamma_{\rm eq}^{\rm max}$. Since $\gamma_{\rm eq}$ is obtained as a linear transformation over $\gamma_e$, its maximum value occurs when $\gamma_e = \gamma_e^{\rm max}$, yielding immediately \eqref{eq:PerfSecCondition}. 
	On the other hand, \eqref{eq:ExpectationNrayPower} is obtained by calculating the expectation of the squared modulus of \eqref{eq:NrayPhysical} and applying the multinomial theorem.  
\end{IEEEproof}

Inspecting \eqref{eq:PerfSecCondition}, we observe that higher values of $R_s$ imply a more restrictive perfect secrecy condition, i.e., if we aim to increase the transmission rate we need $\gamma_b^{\rm min}$ to be larger. This is also shown in Fig. \ref{fig:1}, where increasing $R_s$ shifts $f_{\gamma_{\rm eq}}$ to the right regardless of the considered fading distribution. Moreover, as $\overline{\gamma}_b$ becomes larger -- or, equivalently, $\overline{\gamma}_e$ takes lower values -- we can transmit at a faster secure rate while keeping zero \ac{OPSC}, which is a coherent result. 

We also observe that considering a larger number of rays in \eqref{eq:NrayPhysical} has a significant impact in the \ac{OPSC}. As $N$ increases, either in Bob's or Eve's channel, the interval $[h_k^{\rm min}, h_k^{\rm max}]$ gets wider, causing the condition in \eqref{eq:PerfSecCondition} to be more restrictive. In fact, if $N\rightarrow \infty$, then \eqref{eq:NrayPhysical} becomes \eqref{eq:CLTmodelsPhysical}, implying that $\|h_k^{\rm min}\|\rightarrow 0$ and $\|h_k^{\rm max}\|\rightarrow \infty$, as predicted by \ac{CLT}-based channel modeling approaches.

It is important to note that, although Eve's instantaneous \ac{CSI} is not required, we implicitly assume that the distribution of $h_e$ is known, i.e., the value of $\gamma_e^{\rm max}$, in order to apply the secrecy condition in \eqref{eq:PerfSecCondition}. Because the relative amplitudes of the waves arriving at Eve as well as their average power are closely related to the geometry of the scenario under analysis, this is equivalent to assume that Alice has information over the propagation environment -- or similarly, that is has some sort of statistical knowledge of Eve's \ac{CSI}.

\section{Secure tx over two-wave fading}
\label{S4}
After formulating the conditions on which perfect secrecy can be attained when considering ray-based fading channels, we now analyze a simple albeit rather illustrative case by assuming two dominant components arriving at both receiver ends. The two-wave (or two ray) fading model \cite{Durgin2002,Frolik2007} arises when setting $N=2$ in \eqref{eq:NrayPhysical}, i.e.
\begin{equation}
	\label{eq:2RayPhysical}
	h = V_1 e^{j\phi_1} + V_2 e^{j\phi_2}. 
\end{equation}

This model is completely characterized by the parameter
\begin{equation}
	\Delta = \frac{2V_1V_2}{V_1^2+V_2^2}, 
\end{equation}
which measures the relative difference in amplitude between the two waves. Hence, $\Delta = 1$ implies that both rays have exactly the same power, whilst $\Delta = 0$ signifies that one of the specular components in \eqref{eq:2RayPhysical} vanishes. 

With this consideration, the \ac{PDF} and the \ac{CDF} of the \ac{SNR} at Bob and Eve are written as \cite{Durgin2002,Eggers2019}
\begin{align}
	\label{eq:TwoWavePDF}
	f_{\gamma_k}(\gamma_k) &= \frac{1}{\pi\overline{\gamma}_k\sqrt{\Delta_k^2 - (1-\gamma_k/\overline{\gamma}_k)^2}} & \\
	& &\hspace{-0.5cm}\quad \gamma_k^{\rm min} \leq \gamma_k  \leq \gamma_k^{\rm max}, \notag\\
	F_{\gamma_k}(\gamma_k) &= \frac{1}{2} - \frac{1}{\pi}{\rm asin}\left(\frac{1-\gamma_k/\overline{\gamma}_k}{\Delta_k}\right) \label{eq:TwoWaveCDF}&
\end{align} 

where, as in the previous section, the subindex $k=b,e$ is used to distinguish between the parameters of Bob's and Eve's channel distributions. The domain boundaries for each distribution are calculated as in \eqref{eq:SNRMinMax}, yielding in this case
\begin{align}
\label{eq:avgamma}
	\gamma_k^{\rm min} &= \overline{\gamma}_k (1 - \Delta_k), & \gamma_k^{\rm max} &= \overline{\gamma}_k (1 + \Delta_k),
\end{align}
and therefore the condition for perfect secrecy stated in Proposition \ref{prop:1} is expressed as
\begin{equation}
	\label{eq:SNRconditionTwoWave}
	\overline{\gamma}_b > \frac{2^{R_s}\overline{\gamma}_e(1+\Delta_e) + 2^{R_s}-1}{1-\Delta_b}.
\end{equation}

Thus, despite the fact that Eve's instantaneous \ac{CSI} is unkown at Alice, secrecy in the communication can be ensured if the average \ac{SNR} at Bob is above this threshold. This condition can be met, e.g., by a proper design of the distance between the transmitter and the legitimate receiver, as well as by the definition of secure zones on which no eavesdroppers can be placed. After simple manipulations to \eqref{eq:SNRconditionTwoWave}, the largest constant rate that ensures perfect secrecy is expressed as 
\begin{equation}
	\label{eq:CsTwoWave}
	R_s^{\rm max} = \left[{\rm log}\left(\frac{\overline{\gamma}_b(1-\Delta_b)+1}{\overline{\gamma}_e(1+\Delta_e)+1}\right)\right]^+.
\end{equation}

In fact, whenever Alice has perfect knowledge of Bob's \ac{CSI} (instead of statistical knowledge only), it is possible to adapt its transmission rate to Bob's instantaneous CSI while meeting the condition $\gamma_b > \overline{\gamma}_e(1+\Delta_e)$, which yields the following expression for the instantaneous secrecy capacity:
\begin{equation}
	\label{eq:CsTwoWave2}
	C_s = \left[{\rm log}\left(\frac{{\gamma}_b+1}{\overline{\gamma}_e(1+\Delta_e)+1}\right)\right]^+ \geq R_s^{\rm max}.
\end{equation}

The \ac{OPSC} over two-wave fading is straightforwardly calculated by introducing \eqref{eq:TwoWavePDF} and \eqref{eq:TwoWaveCDF} in \eqref{eq:PoutIntegral}, leading to
\begin{equation}
	\label{eq:OPSCTwoWave}
	P_{\rm out}(R_s) = \frac{1}{\pi\overline{\gamma}_e}\int_{\gamma_e^{\rm min}}^{\gamma_e^{\rm max}} \frac{\widehat{F}_{\gamma_b}\left(2^{R_s}\gamma_e+2^{R_s}-1\right)}{\sqrt{\Delta_e^2- (1-\gamma_e)/\overline{\gamma}_e}}\,d\gamma_e
\end{equation}
with
\begin{equation}
	\widehat{F}_{\gamma_b}(\gamma) = \begin{mycases}
		0, & \text{if}\;\gamma < \gamma_b^{\rm min}\\[1ex]
		\frac{1}{2}-\frac{1}{\pi}{\rm asin}\left(\frac{1-\gamma/\overline{\gamma}_b}{\Delta_b}\right), & \text{if}\;  \gamma_b^{\rm min}<\gamma < \gamma_b^{\rm max}\\[1ex]
		1, & \text{if}\;\gamma > \gamma_b^{\rm min}\\[1ex]
	\end{mycases},
\end{equation}
where the piecewise definition of $\widehat{F}_{\gamma_b}(\gamma)$ is consequence of the finite domain of $F_{\gamma_b}(\gamma_b)$ in \eqref{eq:TwoWaveCDF}. 

\begin{figure}[t]
\centering
\includegraphics[width=\columnwidth]{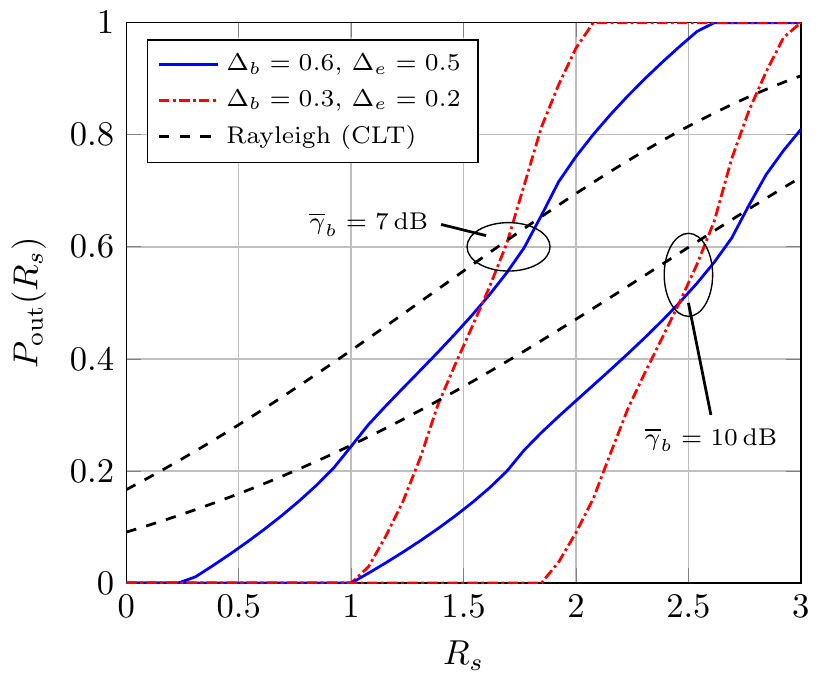}
\caption{Impact of \ac{CLT} based fading models (Rayleigh) and ray-based ones (Two-wave) in the \ac{OPSC} for different values of channel parameters and average \acp{SNR}. For all traces, $\overline{\gamma}_e = 0\,{\rm dB}$.}
\label{fig:2}
\end{figure}

\begin{figure}[t]
\centering
\includegraphics[width=\columnwidth]{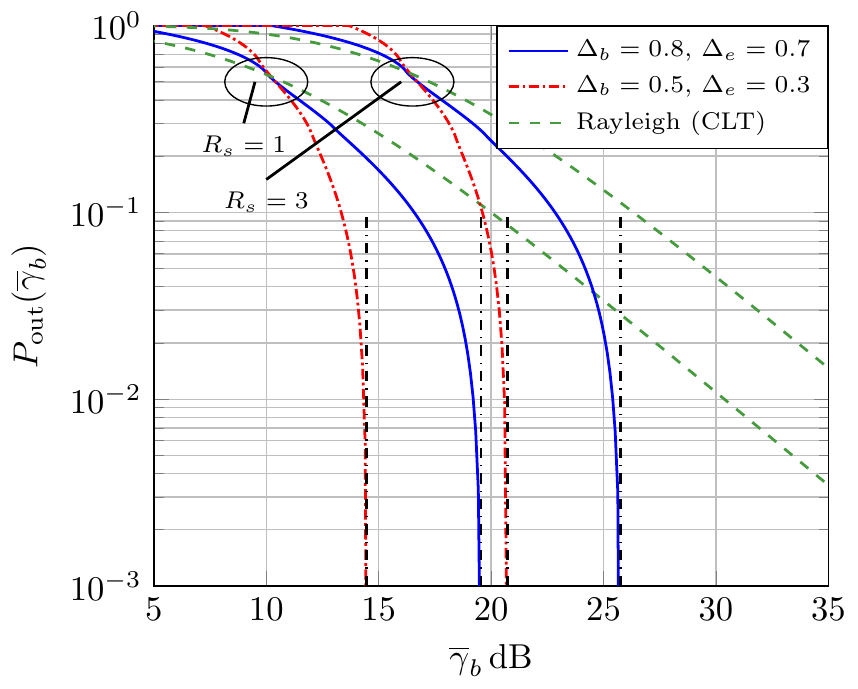}
\caption{\ac{OPSC} in terms of $\overline{\gamma}_b$ for different values of channel parameters and distinct fading models. For all traces $\overline{\gamma}_e = 7\,{\rm dB}$.}
\label{fig:3}
\end{figure}

The outage probability in \eqref{eq:OPSCTwoWave} in terms of $R_s$ and $\overline{\gamma}_b$ is depicted in Figs. \ref{fig:2} and \ref{fig:3}, respectively. For the sake of comparison, $P_{\rm out}$ over \ac{CLT} based channels (in this case, Rayleigh fading) is also shown as a reference. We observe that, for a given $R_s < R_s^{\rm max}$, the outage probability is exactly zero when considering a finite number of reflections, whilst this behavior is not reproduced when assuming a fading model arising from the \ac{CLT}. Specifically, we observe that the asymptotic decay for the Rayleigh case is that of a diversity order equal to one. Conversely, when considering the ray-based alternatives here analyzed the OPSC abrutly drops for the limit value of $\overline\gamma_b$ given by \eqref{eq:avgamma}, which can be regarded as an infinite diversity order. 

We also notice that the parameter $\Delta$ plays an important role in the \ac{OPSC}; as $\Delta_k$ increases, i.e., the power of the rays becomes more similar in either Bob's or Eve's channels, $R_s^{\rm max}$ takes lower values. Then, larger values of $\Delta$ render a lower achievable transmission rate or, equivalently, require higher values of the average \ac{SNR} at Bob for the same $R_s$. It is interesting to pay attention to the limit values of both $\Delta_e$ and $\Delta_b$. While setting $\Delta_e = 1$ still allows to achieve perfect secrecy, substituting $\Delta_b = 1$ in \eqref{eq:CsTwoWave} makes $R_s^{\rm max} = 0$. In the next section, we will see that such restriction vanishes when considering an alternative formulation of \ac{OPSC}. 

\section{Secure and reliable tx over two-ray fading}
\label{S5}
Up to this point, we have considered the classical definition of \ac{OPSC} given in \eqref{eq:PoutDef}. However, this formulation does not distinguish between outage events produced by a failure in achieving perfect secrecy ($R_s > C_s$) or due to the fact that Bob cannot decode the transmitted message (e.g., because its instantaneous \ac{SNR} drops below the minimum value required for a reliable communication) \cite{Zhou2011}. Therefore, we revisit the outage formulation in \cite[Sec. III]{Zhou2011}, according to which the \ac{OPSC} is defined as
\begin{equation}
	\label{eq:PoutdefAlt}
	P_{\rm out}(R_s) \triangleq P\left\{R_s > C_b - C_e \,|\, \gamma_b > \gamma_{\rm th}\right\}
\end{equation}
where $\gamma_{\rm th}\geq 0$ is the minimum \ac{SNR} at Bob required for a reliable communication. Because Bob is supposed to collaborate with Alice, then the latter can suspend the transmission if $\gamma_b < \gamma_{\rm th}$, since it would make no sense transmitting when the legitimate receiver cannot decode the message. With the \ac{OPSC} definition in \eqref{eq:PoutDef}, this situation would produce an outage but, in fact, secrecy is not compromised since there would not be any message transmission. 

Therefore, introducing \eqref{eq:CsDef} in \eqref{eq:PoutdefAlt}, $P_{\rm out}$ is rewritten as 
\begin{equation}
	\label{eq:PoutdefAlt2}
	P_{\rm out}(R_s) = \frac{P\left\{\gamma_{\rm th}<\gamma_b<2^{R_s}\gamma_e + 2^{R_s} -1\right\}}{P\{\gamma_b > \gamma_{\rm th}\}},
\end{equation}
which, after some algebraic manipulations, leads to \eqref{eq:PoutAltIntegral}, placed at the top of next page. We can observe that, if $\gamma_{\rm th}=0$, then \eqref{eq:PoutdefAlt2} becomes \eqref{eq:PoutGameq}, since we eliminate any reliability constraint. 

Consider again the case of a finite number of reflections arriving to the receiver, i.e., the channel gains follow a ray-based distribution as in \eqref{eq:NrayPhysical}. Coming back to the geometrical meaning of the \ac{OPSC}, conditioning $P_{\rm out}$ to the transmission event is equivalent to truncating the left tail of $f_{\gamma_{b}}(\gamma_{b})$ in \mbox{Fig. \ref{fig:1}}. Hence, the perfect secrecy condition is now formulated as
\begin{equation}
	{\rm max}\{\gamma_b^{\rm min},\gamma_{\rm th}\} > 2^{R_s}\gamma_e^{\rm max} + 2^{R_s} -1,
\end{equation}
with $\gamma_b^{\rm min}$ and $\gamma_e^{\rm max}$ given in \eqref{eq:SNRMinMax}. Note that the condition is less restrictive than that in Proposition \ref{prop:1}, allowing us to achieve perfect secrecy in those scenarios where $\gamma_b^{\min}$ takes lower values, i.e.,  $\gamma_b^{\min}\rightarrow 0$. Thus, by properly choosing $\gamma_{\rm th}$, it is possible to ensure secrecy at the expense of a lower transmission probability, which ultimately translates into a reduced throughput. 

\begin{figure}[t]
\centering
\includegraphics[width=\columnwidth]{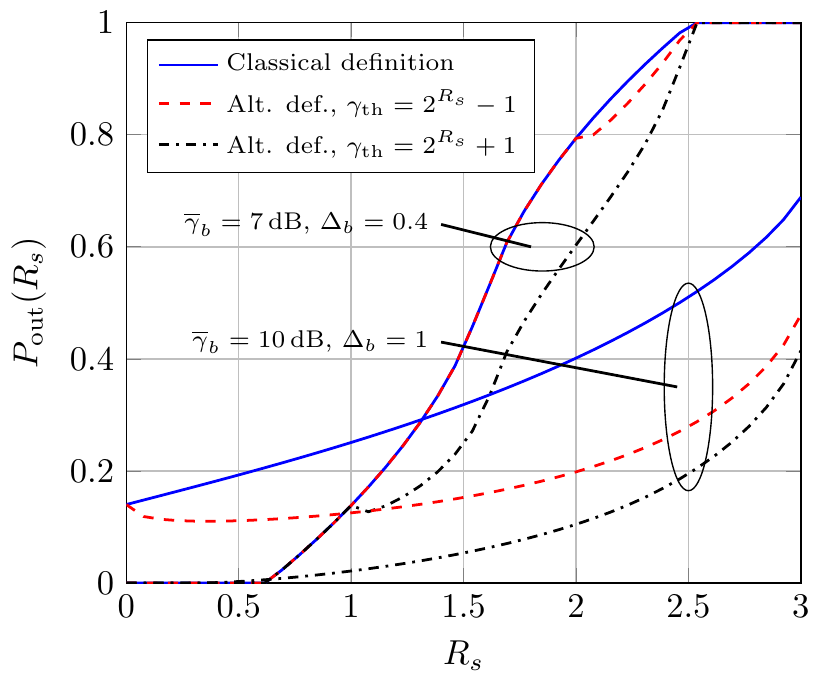}
\caption{Comparison between classical and alternative \ac{OPSC} formulation for different values of channel parameters and distinct \ac{SNR} thresholds. For all traces, $\overline{\gamma}_e = 0\,{\rm dB}$ and $\Delta_e = 0.6$.}
\label{fig:4}
\end{figure}

This is represented in Fig. \ref{fig:4}, where the classical \eqref{eq:PoutDef} and the alternative \eqref{eq:PoutdefAlt} definitions of \ac{OPSC} are compared. The channels gains are assumed to follow a two-wave distribution, and therefore $P_{\rm out}$ is calculated by substituting \eqref{eq:TwoWavePDF} and \eqref{eq:TwoWaveCDF} in \eqref{eq:PoutAltIntegral} and taking into account the boundaries of $F_{\gamma_k}(\gamma_k)$. 
\begin{figure*}[!t]
\setcounter{equation}{26}
\normalsize
\begin{equation}
	\label{eq:PoutAltIntegral}
	P_{\rm out}(R_s) = \frac{1}{1-F_{\gamma_b}(\gamma_{\rm th})}\left[\int_{\left[\frac{\gamma_{\rm th}+1}{2^{R_s}}-1\right]^+}^\infty F_{\gamma_b}\left(2^{R_s}\gamma_e + 2^{R_s}-1\right)f_{\gamma_e}(\gamma_e)\,d\gamma_e \right] - \frac{F_{\gamma_b}(\gamma_{\rm th})\left(1-F_{\gamma_e}\left(\frac{\gamma_{\rm th}+1}{2^{R_s}}-1\right)\right)}{1-F_{\gamma_b}(\gamma_{\rm th})}.
\end{equation}
\hrulefill
\vspace*{4pt}
\end{figure*}
Let us first consider the case on which $\overline{\gamma}_b = 7\,{\rm dB}$ and $\Delta_b = 0.4$. We observe that, until $R_s$ reaches a certain value, $\gamma_{\rm th} < \gamma_b^{\rm min}$ and thus the transmission condition has no impact on the \ac{OPSC}, since it is always met. Naturally, as the threshold increases, such limit value for $R_s$ is reduced.

Regard now the case with $\overline{\gamma}_b = 10\,{\rm dB}$ and $\Delta_b = 1$. As stated before, by choosing a sufficiently large threshold value $\gamma_{\rm th}$, we can ensure perfect secrecy even when $\Delta_b = 1$ (or, equivalently, $\gamma_b^{\rm min} = 0$). However, increasing $\gamma_{\rm th}$ implies a lower throughput, given by $\eta = P\{\gamma_b > \gamma_{\rm th}\} R_s$. 

\section{Impact of the number of scatterers}
\label{S6}
In the previous sections, we have assumed a two-wave distribution for both Bob's and Eve's channel, i.e., $N=2$ in \eqref{eq:NrayPhysical}. Due to the clear impact of $N$ in the perfect secrecy condition stated in Proposition \ref{prop:1}, we are now interested in analyzing the consequences of having a larger number of reflections arriving at the receiver. Specifically, two theoretical scenarios are considered:  (i) \emph{fixed average receive power and different number of scatterers} and (ii) \emph{number of scatterers as a design parameter}.

\subsection{Fixed $\overline\gamma$ and different N}
In this situation, an increased number of reflectors and scatterers renders a richer multipath propagation and, consequently, larger values of both $N_b$ and $N_e$, with $N_b$ and $N_e$ denoting the number of rays in \eqref{eq:NrayPhysical} for Bob's and Eve's channels, respectively. Hence, for some given $\overline\gamma_b$ and $\overline\gamma_e$, our goal is to determine the what extent the consideration of a larger $N_b$ and $N_e$ impacts the secrecy performance. Since the limit case of $\{N_b,N_e\}\rightarrow\infty$ reduces to the Rayleigh fading case, we expect that the perfect secrecy condition in Proposition \ref{prop:1} is not met beyond some limit values of $\{N_b,N_e\}$.

We now express $h_b$ and $h_e$ as
\begin{equation}
	\label{eq:NrayPysicalAlpha}
	h_k = V_{1,k} e^{j\phi_{1,k}} + \sum_{i=2}^{N_k} V_{i,k} e^{j\phi_{i,k}},\quad k = e,b.
\end{equation}
with the amplitudes of the rays given by $V_{i,k} = \alpha_{i,k} V_{1,k}$ for $i=2,\dots,N_k$, with $0<\alpha_{i,k}<1$ and $\alpha_{i,k}\geq\alpha_{j,k}$, $\forall\, i<j$; i.e., the amplitude of the successive rays is expressed as relative to the amplitude of the dominant component.

For simplicity, and to better visualize the impact of increasing $N$, we consider again the classical outage formulation in \eqref{eq:PoutGameq}. Therefore, it is clear that increasing the number of waves at reception makes the secrecy condition more restrictive. On the one hand, if $N_b$ increases, then $\gamma_b^{\rm min}$, which directly depends on $\|h_b^{\rm min}\|$ in \eqref{eq:GainsMinMax}, takes lower values. On the other hand, $\gamma_e^{\rm max}$ also rises with $N_e$.

\begin{figure}[t]
\centering
\includegraphics[width=\columnwidth]{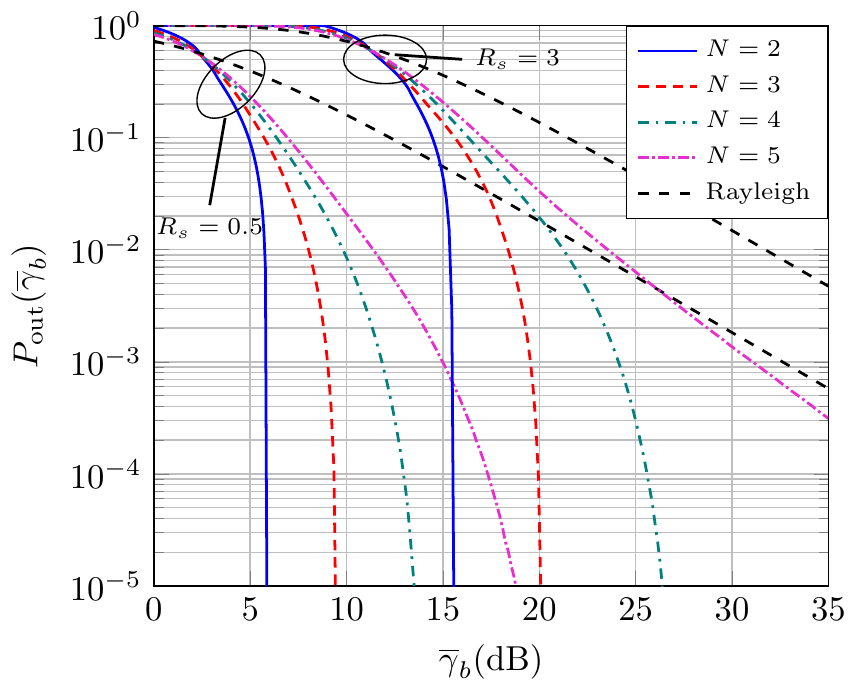}
\caption{Impact of increasing the number of waves at both Eve and Bob. $h_b$ and $h_e$ are distributed according to \eqref{eq:NrayPysicalAlpha} with $\alpha=0.2$ for $R_s = 0.5$ and $\alpha=0.25$ for $R_s = 3$. Also, $N_b = N_e = N$.}
\label{fig:5}
\end{figure}

The effect of increasing the number of rays is studied in Fig. \ref{fig:5}, where the \ac{OPSC} is evaluated for different values of $N=N_b = N_e$. We also set $\alpha_{i,k}=\alpha$, which can be regarded as a worst case situation in terms of secrecy performance.
%
%
Due to the mathematical complexity of the \ac{PDF} of the ray-based model in \eqref{eq:NrayPhysical} when $N>3$, which involves the integral of multiple Bessel's functions \cite{Abdi2000}, we resort on Monte Carlo simulations for this section. 

We observe in Fig. \ref{fig:5} that considering a larger number of waves requires higher values of $\overline{\gamma}_b$ to achieve the same outage probability, for a fixed $R_s$. Moreover, the average \ac{SNR} at Bob needed to ensure perfect secrecy also changes with $N$, which is a coherent result since we are both reducing the value of $\gamma_b^{\rm min}$ and increasing $\gamma_e^{\rm max}$. Note that the relation between the amplitudes $V_i$ also plays a key role on achieving perfect secrecy. For instance, in the case $R_s=3$, in which $\alpha = 0.25$, we cannot ensure a secure transmission for $N=5$, in contrast to the case  $R_s=0.5$ and $\alpha = 0.2$. This is explained as follows: since we need $\gamma_b^{\rm min}>\gamma_e^{\rm max}$, this translates into 
\begin{equation}
	V_{1,b} - \sum_{i=2}^{N_b} V_{i,b} > 0.
\end{equation}

Thus, considering the relation between amplitudes as in \eqref{eq:NrayPysicalAlpha}, we have that $\alpha(N_b-1) < 1$. Hence, if $\alpha = 0.25$ and $N=5$, the condition is not met and therefore no perfect secrecy can be ensured in this case.  

\subsection{Designing N for secrecy}
Let us now move onto the second scenario, on which we assume that we are able to control the number of waves arriving at the receiver ends, i.e., we can somehow \emph{eliminate} some of the rays by a proper design of the propagation characteristics of the environment. This has, obviously, a non-negligible impact on the receiver power, since we are disregarding some components of the channel and hence diminishing its average power. This approach seems desirable for the eavesdropper channel, in the sense that it degrades its average SNR. However, as we will later see, this also turns out being beneficial for the legitimate channel despite the effective decrease on the average SNR at Bob. For this reason, we will specialize our study on the consideration of a fixed number of rays for the eavesdropper channel, and a successive reduction on the number of rays received by Bob.

In order to characterize the \ac{SNR} loss incurred by Bob, we consider the \acp{SNR} at both Bob and Eve given by
\begin{align}
	\gamma_b &= \overline{\gamma}_b \frac{\|h_b\|^2}{\Omega}, & \gamma_e &= \overline{\gamma}_e \frac{\|h_e\|^2}{\Omega},
\end{align} 
where $h_e$ and $h_b$ are given as in \eqref{eq:NrayPhysical} with $N_e = N$ and $N_b < N$, representing the reduced number of waves arriving at Bob. The power loss is characterized by normalizing both channels by\footnote{Note that we are assuming $V_{i,b} = V_{i,e}$ $\forall$ $i$.} $\Omega = \sum_{i=1}^N V_i^2$. Thus, $\mathbb{E}[\|h_b\|^2]/\Omega < 1$, which is equivalent to scale $\overline{\gamma}_b$ by a factor $\mathbb{E}[\|h_b\|^2]/\Omega = \Omega_b/\Omega$. 

\begin{figure}[t]
\centering
\includegraphics[width=\columnwidth]{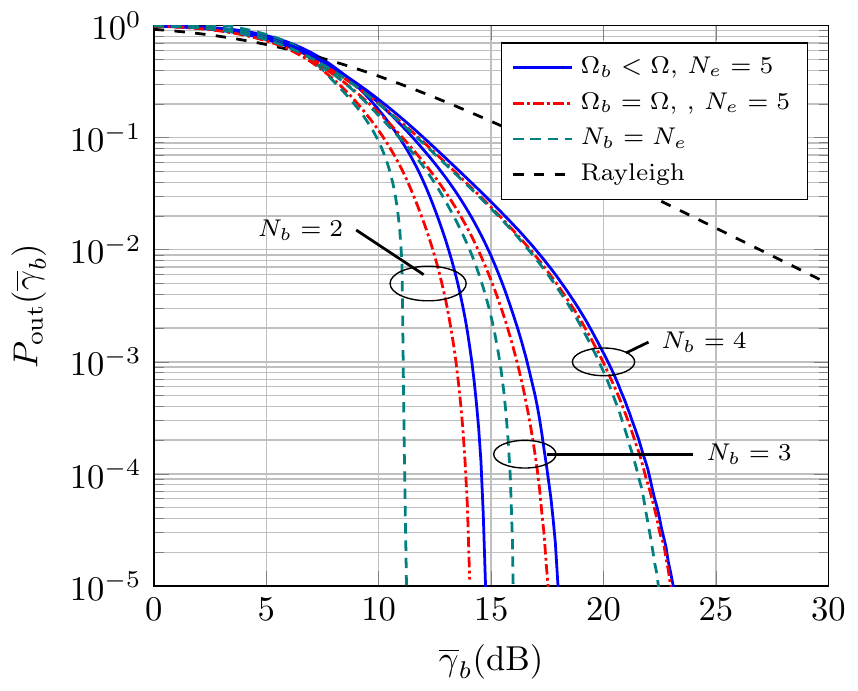}
\caption{Impact of controlling the number of waves arriving at Bob. The wave amplitudes relation is given by \eqref{eq:NrayPysicalAlpha} with $\alpha_{i,k} = \alpha = 0.2$. The case with power loss is compared with the theoretical case in which $\Omega_b = \Omega$ and with that where $N_e = N_b$. For all the traces $R_s$ has been fixed to $R_s = 1$.}
\label{fig:6}
\end{figure}

With this consideration, the \ac{OPSC} is plotted in Fig. \ref{fig:6} for different values of $N_b$ but maintaining the number of waves at Eve. For the sake of comparison, we also include the case of the first scenario in which $N_b = N_e$ and $\Omega_b = \Omega$ and the case  $\Omega_b = \Omega$ but $N_b < N_e$. We observe that, despite the fact that Bob's average \ac{SNR} is lowered, having a reduced number of waves arriving at Bob is beneficial from a secrecy perspective. Note that the impact of eliminating rays on the legitimate channel (and therefore having a lower received power) is less detrimental as $N_b$ approaches $N_e$. In fact, considering $h_k$ as in \eqref{eq:NrayPysicalAlpha} with $\alpha_{i,k} = \alpha$ $\forall$ $i$ and $k = b,e$, the power loss can be written as
\begin{equation}
	\label{eq:PowerRelation}
\frac{\Omega_b}{\Omega}	= \frac{1+\alpha^2(N_b-1)}{1+\alpha^2(N_e-1)}.
\end{equation}

Then, from \eqref{eq:PowerRelation}, it is clear that such loss reduces as $N_b$ increases, being equal to one if $N_b = N_e$, i.e., if we do not \textit{eliminate} any ray. 

Finally, we also note that the most favorable case is that where both Bob and Eve receive a small number of waves, which confirms the beneficial role of a reduced number of scatterers for wireless physical layer security.

\section{Conclusions}
\label{S7}
In this work, we provided a new look at wireless physical layer security, backing off from the classical \ac{CLT} assumption associated to fading and explicitly accounting for the effect of considering a finite number of multipath waves arriving the receiver ends. To the best of our knowledge, we showed for the first time that it is possible to achieve \textit{perfect secrecy} even when the eavesdropper's instantaneous \ac{CSI} is unknown at the transmitter.

We also showed that a rich multipath propagation has a negative impact on the \ac{OPSC}, so that those propagation conditions which imply a reduced number of waves arriving at the receiver ends are instrumental to achieving perfect secrecy. This somehow contradicts the common knowledge that fading is beneficial for physical layer security; this assert is restricted to those situations on which the legitimate channel is more degraded that the eavesdropper's counterpart (and hence \ac{PLS} is not possible in such case in the absence of fading), or when Eve's instantaneous \ac{CSI} is available at Alice.

The consideration of a strong dominant specular component (i.e. larger than the remaining aggregate waves) is the key factor to enable perfect secrecy. Besides, incorporating a reliability constraint in the \ac{OPSC} definition allows for improving the secrecy performance.

Finally, we also pointed out that the ability of controlling the propagation environment in order to reduce the number of waves arriving at the legitimate receiver is also beneficial for physical layer security. This opens up the possibility of using \ac{LIS} to improve secrecy in a complete different way as those suggested in the literature, i.e., to eliminate reflections instead of for maximizing the \ac{SNR} at Bob \cite{Shen2019}.

\bibliographystyle{ieeetr}
\bibliography{bibSecrecyNray}

%
%
%
%

\end{document}